\documentclass[10pt,twocolumn,letterpaper]{article}

\usepackage[pagenumbers]{iccv}   
\usepackage[utf8]{inputenc}     
\usepackage[T1]{fontenc}        

\def\paperID{2512} 
\def\confName{ICCV}
\def\confYear{2025}

\usepackage[T1]{fontenc}    
\usepackage{algorithm}
\usepackage{algorithmic}
\usepackage[table]{xcolor} 
\usepackage{amsfonts}
\usepackage{soul}
\usepackage{amssymb}
\usepackage{booktabs}      
\usepackage{multirow}
\usepackage{ragged2e}
\usepackage{threeparttable}
\usepackage{svg}
\usepackage{stackengine}
\usepackage{bm}
\usepackage{footmisc}
\usepackage{graphicx}
\usepackage{adjustbox}
\usepackage{newfloat}
\usepackage{listings}
\usepackage{amssymb}
\usepackage{pifont}
\usepackage{stackengine}
\usepackage[pagebackref,breaklinks,colorlinks,citecolor=bristol]{hyperref}

\definecolor{cvprblue}{HTML}{118ab2}
\definecolor{teal}{HTML}{047c84}
\definecolor{tiger}{HTML}{fa6a03}
\definecolor{lilac}{HTML}{dc317f}
\definecolor{bristol}{HTML}{ab1f2e}
\definecolor{metablue}{HTML}{0264e0}
\definecolor{darkblue}{HTML}{103c5b}
\hypersetup{
colorlinks=true,
linkcolor=cvprblue,
citecolor=cvprblue,
urlcolor=lilac,
}
\usepackage[most]{tcolorbox}
\tcbset{size=small, frame hidden, colback=white!98!blue!98!green, colframe=white!85!blue!95!green, grow to left by=0.05cm, grow to right by=0.3cm}

\DeclareCaptionStyle{ruled}{labelfont=normalfont,labelsep=colon,strut=off} 
\floatstyle{ruled}
\newfloat{listing}{tb}{lst}{}
\floatname{listing}{Listing}

\newcommand\bmt[1]{\tilde{\bm{#1}}}
\newcommand\bmh[1]{\hat{\bm{#1}}}

\title{GIViC: Generative Implicit Video Compression}
\author{Ge Gao \quad Siyue Teng \quad Tianhao Peng \quad Fan Zhang \quad David Bull  \\ [0.3em]
Visual Information Lab, University of Bristol \\ [0.1em]
{\tt\small \{ge1.gao, siyue.teng, tianhao.peng, Fan.Zhang, Dave.Bull\}@bristol.ac.uk} \\
\small\textbf{\texttt{\url{https://ge1-gao.github.io/GIViC}}}
}

\begin{document}
\maketitle

\begin{abstract}
While video compression based on implicit neural representations (INRs) has recently demonstrated great potential, existing INR-based video codecs still cannot achieve state-of-the-art (SOTA) performance compared to their conventional or autoencoder-based counterparts given the same coding configuration. In this context, we propose a \textbf{G}enerative \textbf{I}mplicit \textbf{Vi}deo \textbf{C}ompression framework, \textbf{GIViC}, aiming at advancing the performance limits of this type of coding methods. GIViC is inspired by the characteristics that INRs share with large language and diffusion models in exploiting \textit{long-term dependencies}. Through the newly designed \textbf{implicit diffusion} process, GIViC performs diffusive sampling across coarse-to-fine spatiotemporal decompositions, gradually progressing from coarser-grained full-sequence diffusion to finer-grained per-token diffusion. A novel \textbf{Hierarchical Gated Linear Attention-based transformer} (HGLA), is also integrated into the framework, which dual-factorizes global dependency modeling along scale and sequential axes. The proposed GIViC model has been benchmarked against SOTA conventional and neural codecs using a Random Access (RA) configuration (YUV 4:2:0, GOPSize=32), and yields BD-rate savings of 15.94\%, 22.46\% and 8.52\% over VVC VTM, DCVC-FM and NVRC, respectively, on the UVG test set. As far as we are aware, GIViC is the first INR-based video codec that outperforms VTM based on the RA coding configuration. The source code will be made available.
\end{abstract}

\section{Introduction} 
\label{sec:introduction}
The ubiquitous consumer demand for high-quality digital video has accelerated the development of increasingly powerful compression techniques~\cite{bull2021intelligent}. While the latest video standards, such as MPEG H.266/VVC~\cite{bross2021overview} and AOM (Alliance for Open Media) AV1~\cite{han2021technical}, offer impressive coding efficiency and architectural compatibility with previous standards, their coding gains are achieved through the use of increasingly sophisticated tools built upon the conventional hybrid video coding framework. In contrast, neural video compression~\cite{lu2019dvc,li2021deep,li2024neural} has emerged in recent years as a data-driven framework, leveraging end-to-end optimization to achieve a performance level that rivals, or in some cases, surpasses~\cite{xiang2022mimt,li2024neural,jiang2024ecvc} that of standard video codecs. 

More recently, implicit neural representation (INR) based solutions~\cite{chen2021nerv} have provided a more flexible, and potentially lightweight, alternative to these `generic' neural video codecs. By adaptively \textit{overfitting} a neural network to a specific (input) video sequence, INR-based video codecs~\cite{lee2023ffnerv,chen2023hnerv,kwan2024hinerv} exploit long-term spatiotemporal dependencies through sequence-level parameter sharing and stochastic optimization, showing the potential to achieve competitive coding performance~\cite{gao2024pnvc,kwan2024nvrc}.

\begin{figure}[t]
    \centering
    \adjustbox{left=4.7cm, right=0.4cm}{
        \includegraphics[width=1\linewidth]{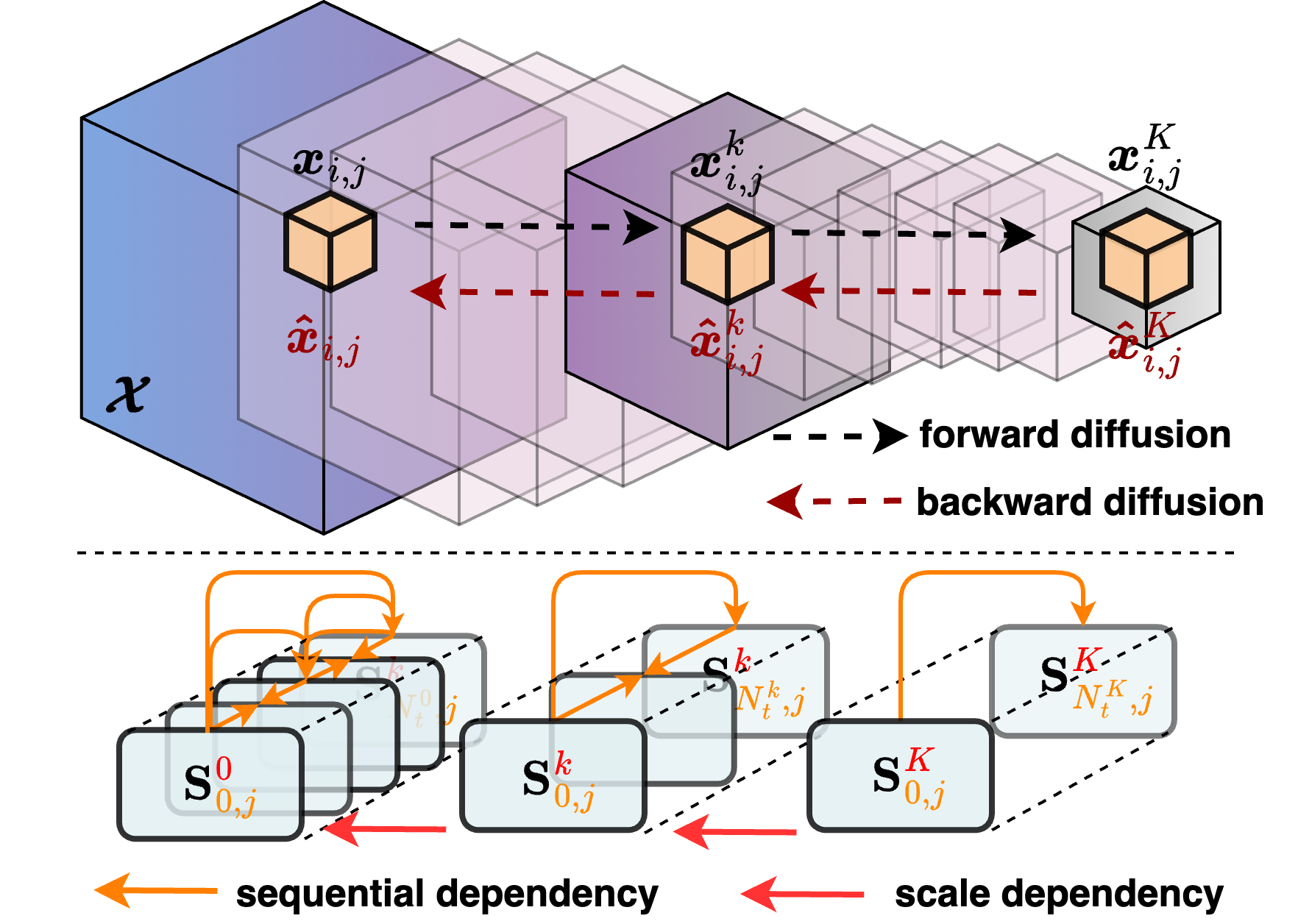}
    }
    \caption{(\textbf{Top}) Illustration of the implicit diffusion framework based on spatiotemporal downsampling of a GOP $\bm{\mathcal{X}}$ with additive noise, interlinking independent diffusion within constant-sized tokens $\{\bm{x}^k_{i,j}\}$ across $k = 1, \dots, K$ levels of abstractions.  (\textbf{Bottom}) The global spatiotemporal dependencies are captured by the 2D hidden states $\mathbf{S}^{\textcolor{red}{k}}_{\textcolor{orange}{{i,j}}}$ of the HGLA transformer, recurrently updated along both \textcolor{red}{scale} and \textcolor{orange}{sequence} axes.}
    \label{fig:illustration-of-GIViC}
\end{figure}

However, the application scenarios, and more critically, the compression performance of existing INR-based codecs are generally limited by their encoding latency, i.e., the number of consecutive frames that can be represented with a single set of learnable parameters. When the system latency is constrained to be compatible with the Low Delay or Random Access configurations~\cite{vtm_ctc} typically used in standard video codecs, INR-based methods~\cite{gao2024pnvc} are outperformed by SOTA conventional codecs such as VVC VTM~\cite{vtm} and generic neural codecs such as the recently improved DCVC models~\cite{li2024neural,qi2025long}.

In this paper, we enhance the INR framework with powerful diffusion models (DMs) and transformer backbones, sufficiently capacitated to model \textit{full-GOP-level spatiotemporal dynamics}. The resulting video compression framework, GIViC (\textbf{G}enerative \textbf{I}mplicit \textbf{Vi}deo \textbf{C}ompression), is built on a novel conditional implicit diffusion model, as shown in \autoref{fig:illustration-of-GIViC} (top). This decomposes a joint diffusion process into cascaded spatiotemporal pyramids, where each stage is extrapolated from denoised representations at coarser scales and previously denoised reference tokens, accelerating denoising while preserving representation quality. GIViC also integrates HGLA (Hierarchical Gated Linear Attention), a novel linear transformer that harmonizes efficiency and effectiveness by dual-factorizing long-term dependency modeling along both scale and sequence axes. Leveraging hierarchically gated recurrence, HGLA scales linearly with long context length spanning the entire GOP, as illustrated in \autoref{fig:illustration-of-GIViC} (bottom). The main contributions of this paper are summarized as follows:

\begin{itemize}
\item This is the \textbf{first time} diffusion models and transformers have been jointly integrated into an INR-based framework for video compression, resulting in a highly expressive architecture for full-GOP-level distribution modeling that enables SOTA compression performance.

\item We propose a novel and integrated \textbf{implicit diffusion framework} for video coding that decomposes the diffusion process into spatiotemporal pyramidal stages, interlinking coarser-grained global variations with finer-grained local details during the per-token diffusion denoising process. 

\item We also introduce a \textbf{new linear transformer architecture}, \textbf{HGLA}, that captures long-term dependencies jointly along scale and sequence axes. HGLA achieves linear complexity w.r.t (long) context lengths while maintaining competitive performance compared to vanilla transformers that have quadratic complexity.

\end{itemize}

We have benchmarked GIViC against SOTA conventional and neural video codecs on the UVG, MCL-JCV, and JVET-B datasets under the Random Access (RA) configuration (YUV colorspace). Results demonstrate significant coding gains, with GIViC outperforming VTM 20.0, DCVC-FM, and NVRC by 15.94\%, 22.46\%, and 8.52\%, respectively, on the UVG test set, and by 7.71\%, 22.34\%, and 16.13\%, respectively, on the JVET-B test set. To the best of our knowledge, GIViC is the \textbf{first INR-based video codec to surpass VTM performance} in the RA coding mode.

\section{Related Work} \label{sec:realted-work}

\noindent \textbf{Neural video compression.} The focus of video compression research is progressively shifting from conventional hand-crafted codecs~\cite{wiegand2003overview,sullivan2012overview,bross2021overview} to those that incorporate learning-based enhancement of individual coding tools~\cite{afonso2018video,zhang2021video,zhang2021video} often within end-to-end optimized coding frameworks.  Recent contributions have been based on various innovations including: improving sub-components~\cite{hu2021fvc,mentzer2022vct,xiang2022mimt,ho2022canf,li2023neural,jiang2024ecvc}, optimizing rate control~\cite{xu2023bit,zhang2024neural}, leveraging instance-specific overfitting~\cite{van2021overfitting,oh2024parameter} and accelerating inference~\cite{hu2023complexity,peng2023accelerating}. Currently, the best-performing model~\cite{qi2025long} has been reported~\cite{teng2024benchmarking} to offer improved performance over ECM~\cite{ECM_GitLab} under the Low-Delay configuration.

\vspace{5pt}

\noindent \textbf{Implicit neural representations (INRs)}, which use neural networks to map multimedia signals into coordinate-based representations~\cite{sitzmann2020implicit,dupont2021coin,xu2022signal,jiang2024hiif}, offer an efficient and elegant (albeit unconventional) alternative for video compression. INR-based methods~\cite{chen2021nerv,zhao2023dnerv,lee2023ffnerv,chen2023hnerv,he2023towards,bai2023ps,yan2024ds} exploit sequence-level spatiotemporal redundancy by encoding video sequences within a compact set of network parameters, reformulating visual data compression into a model compression task that leverages pruning, quantization, and entropy penalization techniques~\cite{han2015deep,gomes2023video,zhang2024boosting}. While recent advances have improved the compression efficiency of INRs through hierarchical encoding~\cite{kwan2024hinerv} and more advanced compression methods~\cite{kwan2024nvrc,gao2024pnvc}, they are still outperformed by SOTA conventional~\cite{vtm,ECM_GitLab} and neural video codecs~\cite{li2024neural,qi2025long} under the same latency constraints.

\vspace{5pt}

\noindent \textbf{Long sequence modeling.} Recently, the success of Large Language Models (LLMs)~\cite{deletang2024language,huang2024compression} has been driven by a core principle in information theory - \textit{jointly modeling long token sequences can maximize compression efficiency}~\cite{thomas2006elements}. However, unlike natural languages, visual signals are associated with bidirectional dependencies that defy simple unidirectional structures, resulting in poorer performance with decoder-only architectures compared to diffusion and non-autoregressive methods~\cite{sun2024autoregressive,wang2024emu3}. Additionally, the quadratic complexity of self-attention in LLMs poses challenges for scaling to long contexts, motivating the development of linear attention~\cite{yang2023gated,peng2023rwkv,gu2023mamba}, which enables parallelized training, linear complexity inference, and performance comparable to standard transformers. This design has been recently adopted in some neural compression methods~\cite{feng2025linear,qin2024mambavc,jiang2024ecvc}.

\begin{figure*}[!t]
\centering
    \includegraphics[width=1\linewidth]{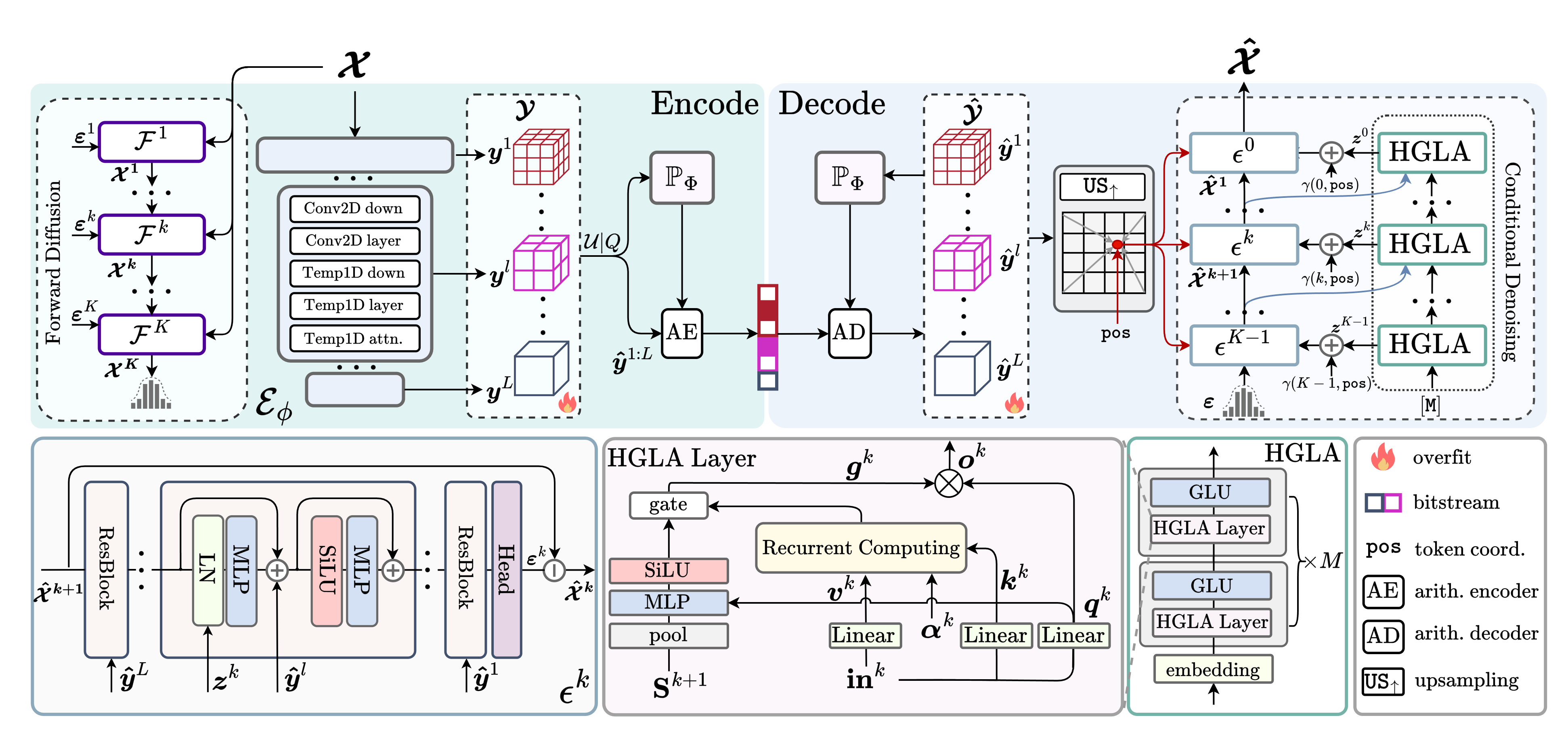}
    \caption{Illustration of the GIViC network architecture.}
    \label{fig:framework}
\end{figure*}

\vspace{5pt}
\noindent \textbf{Diffusion models.} Diffusion models (DMs)~\cite{ho2020denoising,song2021denoising,danier2024ldmvfi} have proved to be more reliable and expressive compared to other types of generative models, e.g., VAEs~\cite{kingma2013auto} and GANs~\cite{goodfellow2014generative}. They have contributed to (generative) visual compression by unconditionally communicating lossy Gaussian samples~\cite{theis2022lossy} or by generating photorealistic images conditioned on entropy-encoded information~\cite{yang2024lossy,relic2024lossy,zhou2024controllable,zhang2024video}. Although most diffusion models enforce a fixed forward corruption process and operate at a single resolution~\cite{song2021denoising}, recent studies~\cite{ho2022cascaded,teng2023relay,bansal2024cold,daras2023soft} demonstrate a more generalized and efficient alternative, i.e., performing diffusion across multiple resolutions and incorporating \textit{arbitrary} degradations such as blurring and vector quantization.

\section{Methods} \label{sec:methods}

Let $\bm{\mathcal{X}} \in \mathbb{R}^{T \times H \times W \times 3}$ be a GOP (Group of Pictures) with $T$ consecutive video frames with spatial resolution $H \times W$. As shown in \autoref{fig:framework}, a set of latents $\bm{\mathcal{Y}} = \{ \bm{y}^l \}^L_{l=1}, \bm{y}^l \in \mathbb{R}^{S^l_t \times S^l_h \times S^l_w \times D^l}$, embedded within a compact local grid structure~\cite{kwan2024hinerv}, is initialized by a spatiotemporal encoder $\mathcal{\bm{E}}_\phi$, i.e., $\bm{\mathcal{Y}} = \mathcal{\bm{E}}_\phi(\bm{\mathcal{X}})$ and stochastically overfitted to $\bm{\mathcal{X}}$ during encoding. Here $(S^l_t, S^l_h, S^l_w)$ denotes the local grid size at level $l$, and $D^l$ is the channel dimension. The complete representation $\bm{\mathcal{Y}}$ is quantized into $\bmh{\mathcal{Y}} = \{ \bmh{y}^l \}^L_{l=1}$ by $Q(\cdot)$, which we relax as $q_\phi(\bm{\mathcal{Y}}|\bm{\mathcal{X}})=\mathcal{U}(\bm{\mathcal{Y}}-0.5, \bm{\mathcal{Y}} + 0.5)$ during encoding to address the non-differentiability issue~\cite{balle2018variational}. A context model $\mathbb{P}_\Phi(\cdot)$ is then used to evaluate the probability mass function (PMF) of $\bmh{\mathcal{Y}}$, with which $\bmh{\mathcal{Y}}$ could be losslessly entropy encoded into the bitstream. 

At the decoder, the quantized hierarchical latents $\bmh{\mathcal{Y}}$ are entropy decoded from the bitstream based on the same context model $\mathbb{P}_\Phi(\cdot)$, which is recurrently updated by the previously decoded latents from the previous spatiotemporal \textit{subgroups} - see the description in \textbf{Tokenization and shuffling} for detailed definitions. $\bmh{\mathcal{Y}}$ contains per-token visual priors that could be extracted via coordinate-based interpolation and used to steer the denoising process towards faithfully reconstructing $\bmh{\mathcal{X}}$. The conditional denoising processing is based on a \textit{per-token} denoising diffusion variational autoencoder, implemented by unrolling a small, $L$-layer (plus a head layer mapping to the pixel space) INR-based denoiser over $K$ representation levels, i.e., $\bm{\epsilon}_\theta := \{ \epsilon^k\}^{K-1}_{k=0}$. Each $\epsilon^k$ predicts the noise $\bm{\varepsilon}^k$ at step $k$, from which the denoised output is produced as $\bmh{\mathcal{X}}^{k}$, conditioned by $\gamma(k, \texttt{pos}) + \bm{z}^k$. Here $\gamma(\cdot)$ denotes positional embedding~\cite{su2024roformer}, $\texttt{pos}$ denotes the token's global 3D coordinate, and $\bm{\mathcal{Z}}^k$ is the conditioning vector sampled by the HGLA transformer based on the previously denoised output $\bmh{\mathcal{X}}^{k+1}$ (or a mask token $[\texttt{M}]$ if $k=K$) as input, and its parameters are optimized offline.

\subsection{Spatiotemporal Encoder}
The spatiotemporal encoder $\mathcal{\bm{E}}_\phi$ relies on large spatiotemporal receptive fields to generate consistent and compact latents. To avoid pretraining $\mathcal{\bm{E}}_\phi$ entirely from scratch, we instead `inflate' a pretrained image autoencoder~\cite{duan2023lossy} to handle the additional temporal dimension, inspired by recent image-to-video generation methods~\cite{blattmann2023align,yu2024efficient}, by inserting temporal downsampling, convolutional, and attention layers in between 2D spatial operations, as illustrated in~\autoref{fig:framework}.

\subsection{Implicit diffusion} \label{sec:implicit-diffusion}
The proposed implicit diffusion is a generalized~\cite{rissanen2023generative,bansal2024cold,gu2023fdm} 
spatiotemporal pyramidal framework, which subsumes traditional explicit multi‐resolution diffusion models into a single, continuous forward‐reverse chain. Instead of treating sub-band and resolution transitions separately, we embed them implicitly within a unified process, ensuring end-to-end consistency across continuous representation scales. We perform per-token diffusion while maintaining a constant token size across scales, eliminating the need to chain separate models at progressively higher resolutions. This design significantly enhances training and inference efficiency by leveraging \textbf{(i)} distributed computations across multiple spatiotemporal resolutions, and \textbf{(ii)} a novel dual-factorized conditional denoising strategy which enables finer-grained subspaces to be extrapolated from those at coarser scales and fully decoded/denoised reference frames. 

\vspace{5pt}
\noindent \textbf{Tokenization and shuffling.} We first define a \textit{tokenization-and-shuffle} operation that specifies the order by which tokens are denoised. We first partition the input GOP $\bm{\mathcal{X}}$ into $N_t = \lceil T/r_t \rceil$ subgroups along the temporal dimension and reorder these subgroups according to the hierarchical frame structure in the RA configuration used by modern standard video codecs~\cite{vtm_ctc}. Here $r_t$ is the resolution of the token in the temporal domain. Within each subgroup (indexed by $i$), the frames are further patchified into continuous-valued, non-overlapping 3D tokens with spatial resolution $(r_h, r_w)$, yielding $\{ \bm{x}_{i,j} \}^{N_s}_{j=1}$, where $N_s = \lceil H/r_h \rceil \times \lceil W/r_w \rceil$ and $\bm{x}_{i,j} \in \mathbb{R}^{r_t \times r_h \times r_w \times 3}$. The partitioned tokens are further grouped spatially according to the Quincunx pattern~\cite{el2022image}, resulting in $\{ \bm{x}_{i,j}\}_{j \in G_d},d=1,\dots, 5$. This defines the (causal) order in which tokens are decoded. Here $d$ stands for the spatial decoding step, $G_d$ denotes the group of tokens in this temporal subgroup ($i$) that are decoded at step $d$, and $|G_1|=|G_2|=\lceil N_s/16 \rceil$ and $|G_d| = 2 \times |G_{d-1}|$ for $d = 3,4,5$. Based on this grouping method, the number of tokens decoded per step is \textbf{doubled} along both spatial and temporal axes, which reduces the number of decoding steps\footnote{Here, we assume $\log_2(N_t -1) \in \mathbb{Z}$.} to $5K \cdot \left(\log_2\left(N_t -1\right)+2\right)$.

\vspace{5pt}
\noindent \textbf{Forward diffusion.} We define a forward diffusion process that entails a sequence of transforms $\bm{\mathcal{F}} = \{ \mathcal{F}^k \}^K_{k=1}$ that progressively ``corrupt'' $\bm{\mathcal{X}}$:
\begin{align}\label{eq:forward-corruption}
&\bm{\mathcal{X}}^k = \mathcal{F}^{k}(\bm{\mathcal{X}},\bm{\varepsilon}^k) = \texttt{DS}\bigl( \bm{\mathcal{X}}, \bm{R}^k \bigr) + \bar{\beta}^k \bm{\varepsilon}^k, 
\end{align}
in which $\bm{\mathcal{X}}^k \in \mathbb{R}^{T^k \times H^k \times W^k \times 3}$, $T^k = \lceil T/R^{k}_t \rceil $, $H^k = \lceil H/R^{k}_s \rceil $, and $W^k = \lceil W/R^{k}_s \rceil$. $\bm{\varepsilon}^k \sim q(\bm{\varepsilon}^k)$ represents the Gaussian noise with a normal distribution. $\texttt{DS}(\cdot, \bm{R}^k)$ denotes the trilinear downsampling operation that downsamples the input temporally and spatially by a factor of $\bm{R}^k=(R^k_t, R^k_s)$ at scale $k$. $q(\cdot)$ corresponds to a normal distribution. When $k=1$, $\bm{\mathcal{X}}^{0}\equiv \bm{\mathcal{X}}$. $\bar{\beta}^k$ is the noise scheduling parameter controlling the strength of noise at step $k$.

Tokenization-and-shuffle is also applied to each $\bm{\mathcal{X}}^k$, producing re-ordered 3D tokens $\{ \{ \bm{x}^k_{i, j} \}_{j\in G^k_d} \}^{N^k_t}_{i=1}$ with the constant token size $(r_t, r_s, r_s)$, where $N_t^k = \lceil T^k / r_t \rceil$ and $N_s^k = \lceil H^k/r_s\rceil \times \lceil W^k/r_s\rceil$. In this way, the number of tokens is reduced with the scales and each token $\bm{x}^k_{i, j}$ encapsulates the visual contents of multiple corresponding tokens in $\bm{\mathcal{X}}^{k-1}$ at a coarser scale. The transforms $\mathcal{F}^1 \cdots \mathcal{F}^k$ gradually destroy the original information, i.e., $\mathbb{I}(\bm{\mathcal{X}}, \bm{\mathcal{X}}^k) \leq \mathbb{I}(\bm{\mathcal{X}}, \bm{\mathcal{X}}^{k-1}), \forall k \in \{1, \dots, K\}$ and $\mathbb{I}(\bm{\mathcal{X}}, \bm{\mathcal{X}}^K) \approx 0$. $\mathbb{I}(\cdot, \cdot)$ denotes the mutual information.

\vspace{5pt}
\begin{figure}
    \centering
        \includegraphics[width=\linewidth]{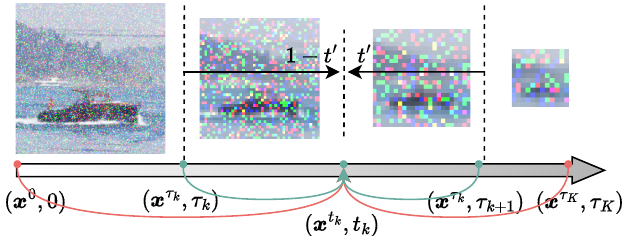}
    \caption{Illustration of cross-resolution consistency training.}
    \label{fig:diffusion-interp}
\end{figure}

\noindent \textbf{Conditional denoising.} 
At step $k$ (with the corresponding denoiser $\epsilon^k$), the per-token denoised output\footnote{We omit the subscripts $i,j$ in~\autoref{sec:implicit-diffusion} for simplicity.} $\bmh{x}^{k-1}$ is derived by:
\begin{align}
    \bmh{x}^{k-1} &= \bmh{x}^k - \bar{\beta}^k\bm{\varepsilon}^k \notag \\
    &= \bmh{x}^k - \bar{\beta}^k\epsilon^k\left(\bmh{x}^k; \bm{z}^k+\gamma(k,\texttt{pos}), \{\bmh{y}^l\}^L_{l=1}\right),
\end{align} 
where $\bm{z}^k$ is the corresponding conditioning token of $\bm{x}^k$ produced by the HGLA transformer ($\bm{\mathcal{Z}} = \{ \bm{z}^k \}$). $\gamma(\cdot)$ denotes the positional embedding~\cite{su2024roformer}.

We then define a set of uniform intervals \( 0 = \tau_0 < \tau_1 < \dots < \tau_K = 1 \), as shown in~\autoref{fig:diffusion-interp}, along the continuum of spatiotemporal resolutions, which partitions the diffusion time interval \([0, 1]\) into \( K \) sub-intervals. Here, we allow \( K \rightarrow \infty \) during training. Considering a random diffusion step $t_k$ falling into the sub-interval \([\tau_k, \tau_{k+1})\), we calculate the normalized position of $t_k$ within the interval as $t' = (t_k-\tau_k)/(\tau_{k+1}-\tau_k)$. The corresponding denoised token $\bmh{x}^{t_k}$ could be yielded via two interpolation paths ``blending'' the higher-resolution clean and lower-resolution noisy counterparts:
\begin{align}
    \bmh{x}_{(1)}^{t_k} =& t_k \texttt{DS}(\bm{x}, \bm{R}^{t_k}) + (1 - t_k) \texttt{US}(\bm{x}^K, \bm{R}^{\tau_k}/{\bm{R}^{t_K}}), \\
    \notag \bmh{x}_{(2)}^{t_k} =& t' \texttt{US}(\bm{x}^{\tau_{k+1}}, {\bm{R}^{t_k}}/{\bm{R}^{\tau_{k+1}}}) \\
    &\ \ \ \ \ \ \ \ \ \ \ \ \ \ \ \ \ \ \ \ \ \
    + (1-t') \texttt{DS}(\bm{x}^{\tau_k}, {\bm{R}^{\tau_k}}/{\bm{R}^{t_k}}),
\end{align}
where $\texttt{US}(\cdot, \cdot)$ stands for the upsampling operation, similar to $\texttt{DS}$. With this formulation, the $K$ stages at inference time could be viewed as a discretization of continuous, densely sampled downsampling stages over timesteps $\tau \in [0, 1]$ during training. The denoising objective $\mathcal{L}_{\text{consistency}}(\theta)$ is defined as:
\begin{align} \label{eq:consistency-obj}
    \mathcal{L}_{\text{consistency}}(\theta) = \mathbb{E}\left[ \| \bm{ \epsilon}_\theta(\bmh{x}^{t_k}_{(1)}, t^k) - \bm{\epsilon}_\theta(\bmh{x}^{t_k}_{(2)}, t^k) \|^2 \right],
\end{align}
which enforces that the output at time $t^k$ for arbitrary $k$ is similar regardless of the path taken.

\subsection{HGLA Transformer} \label{sec:HGLA}
We propose a linear attention based transformer~\cite{katharopoulos2020transformers,peng2023rwkv,gu2023mamba,yang2023gated} dubbed HGLA (Hierarchically Gated Linear Attention), which leverages fixed-size 2D hidden states to store historical contexts, enabling recurrent updates that are parallel during training and of linear complexity w.r.t context length during inference. HGLA\footnote{In~\autoref{sec:HGLA} we use the notation for the case of diffusion conditioning (i.e., $k = 1, \dots, K$), however the idea is easily generalizable (as used for the entropy model $\mathbb{P}_\Phi$).} is the backbone in our framework for denoising conditioning, i.e., $\bm{\mathcal{Z}}^{k} \sim \prod_{k=K}^{k} p_{\psi}(\bm{\mathcal{Z}}^{k}|\mathcal{\hat{\bm{X}}}^{K:k+1})$, and for modeling the hierarchical prior $\prod_{l=L}^{1} \mathbb{P}_{\Phi}(\bm{\mathcal{Y}}^l)$. It is inspired by HGRN2~\cite{qin2024hgrn} and tailored for the GIViC's multi-scale design by maintaining matrix-valued state $\mathbf{S}^k_{i,j}$ per representation level $k$ that is updated per spatiotemporal subgroup for scale $k$ and interacts across scales, which facilitates long-term dependency modeling along both scale and sequence axes.

\vspace{5pt}
\noindent \textbf{Architecture.} Each HGLA module comprises $M$ HGLA layers (as shown in~\autoref{fig:framework}). The dynamics of $\mathbf{S}^k_{i,j}$ are modulated by data-dependent decays, based on cumulative softmaxing \texttt{cumax}~\cite{qin2024hgrn} along the sequence dimension, that specify a lower bound $\bm{\alpha} \in \mathbb{R}^{M \times h}$ on how rapidly the historical contexts are updated, thus guiding lower and upper layers of each HGLA module to focus on short- and long-term dependencies, respectively. Specifically, we maintain a set of $\bm{\Gamma}^k \in \mathbb{R}^{M \times h}$ per representation scale $k$, where $h$ stands for the hidden state dimension, and introduce a bias term $\Delta^k=\log (\frac{R^k_t \cdot R^k_s}{R^K_t \cdot R^K_s}) \cdot \mathbf{1}_{M \times h}$, resulting in the gating $\bm{\alpha}^k \in \mathbb{R}^{M \times h}$:
\begin{gather}
\bm{\alpha}^k = \texttt{cumsum}\left(\texttt{softmax}\left(\bmt{\Gamma}^k, \text{dim}=0 \right), \text{dim}=0 \right), \notag \\
\text{where} \quad \bmt{\Gamma}^k = \bm{\Gamma}^k + \Delta^k.
\end{gather}
Further, to support cross-scale information propagation, we attend the query $\bm{q}_{i,j}^k$ to a \textit{learned mixture} of the current scale's hidden state $\mathbf{S}^k_{i,j}$ and the \textit{updated} hidden state from the coarser scale $\mathbf{S}^{k+1}_{\prec(i,j)}$, based on the gating value $\bm{g}^k_{\prec(i,j)} \in \mathbb{R}^h$, where we use $\prec(i,j)$ to denote token indices that are smaller than $i$ spatially and $j$ temporally\footnote{We have $\mathbf{S}^k_{i,j} \equiv \bm{S}^k_{\prec (i',j')}$ where $i', j'$ entail the indices for all tokens to be decoded the next step.}. The aforementioned HGLA operation at the scale $k$ is formalized as:
\begin{gather}
\bm{g}^k_{i, j} = \sigma\left( \bm{W}_{\bm{g}} \left( [\bm{q}^k_{i, j}; \text{pool}\left( 
\mathbf{S}^{k+1}_{i, j} \right) ]\right) + \bm{b}_{\bm{g}} \right), \\
\mathbf{S}^k_{i, j} = \mathbf{S}^{k+1}_{i,j} \cdot \text{Diag}(\bm{\alpha}^k_{i,j}) + \sum \bm{v}^k_{i,j} \otimes \bm{k}^k_{i,j}, \\
\bm{o}^{k}_{i,j} = \left(\bm{g}^{k}_{i,j} \mathbf{S}^{k+1}_{i,j} + \left(\mathbf{1} - \bm{g}^{k}_{i,j} \right) \mathbf{S}^{k}_{\prec (i,j)} \right) \cdot \bm{q}^k_{i,j},
\end{gather}
where $\text{pool}(\cdot)$ denotes the average pooling operation and $\bm{o}^k_{i, \forall j}$ is the output of the HGLA module. 
Here, the queries $\bm{q}^k_{i,j}$, keys $\bm{k}^k_{i,j}$, and values $\bm{v}^k_{i,j}$ for the module's input $\mathbf{in}^k_{i,j} \in \mathbb{R}^h$ are generated as,
\begin{equation}
    \bm{q}^k_{i,j} = \bm{W}_{\bm{q}} \mathbf{in}^k_{i,j}, \bm{k}^k_{i,j} = \bm{W}_{\bm{k}} \mathbf{in}^k_{i,j}, \bm{v}^k_{i,j} = \bm{W}_{\bm{v}} \mathbf{in}^k_{i,j},
\end{equation}
in which $\{ \bm{W}_{\bm{q}}, \bm{W}_{\bm{k}}, \bm{W}_{\bm{v}} \in \mathbb{R}^{h \times h} \}$ are the learnable linear embedding matrices. It is noted that, for the initial decoding step where the first group of $\bmh{x}^K_{i,j}$ (or, equivalently, $\bmh{y}^L$ in the case of context modeling) has not been decoded, we feed the transformer(s) with a set of learned mask tokens $[\texttt{M}]$.

\begin{table*}[!t]
\centering
\resizebox{\textwidth}{!}{
\begin{tabular}{r|rr|rr|rr|rr|c|c}
\toprule[1.5pt]
BD-rate (\%) & \multicolumn{2}{c|}{UVG} & \multicolumn{2}{c|}{MCL-JCV} & \multicolumn{2}{c|}{JVET-B} & \multicolumn{4}{c}{model complexity} \\
\midrule[1.1pt]
codec & $\text{PSNR}$ & $\text{MS-SSIM}$ & $\text{PSNR}$ & $\text{MS-SSIM}$ & $\text{PSNR}$ & $\text{MS-SSIM}$ & enc. FPS & dec. FPS & params (M) & kMACs/px \\
\midrule
HM 18.0 (RA)~\cite{hm}              & -45.65 & -40.67 & -44.41 & -40.01 & -43.88 & -46.94 & 0.06 & 39.5 & N/A & N/A  \\
VTM 20.0 (RA)~\cite{vtm}            & -15.94 & -16.19 & -11.44 & -8.57 & -7.71 & -6.32 & 0.02 & 23.1 & N/A & N/A \\
\midrule
VCT~\cite{mentzer2022vct}           & -58.32 & -59.21 & -57.01 & -57.92 & -60.21 & -62.27 & 2.21 & 0.69 & 154.2 & 5536 \\
DCVC-DC~\cite{li2023neural}         & -36.68 & -40.12 & -37.03 & -31.13 & -36.42 & -30.63 & 0.99 & 1.39 & 50.8 & 1274 \\
DCVC-FM~\cite{li2023neural}         & -22.46 & -27.23 & -20.31 & -21.01 & -22.34 & -24.28 & 0.93 & 4.87 & 44.9 & 1073 \\
\midrule
C3 (adaptive)~\cite{kim2024c3}      & -60.21 & -58.99 & -56.01 & -55.81 & -- & -- & 3.21 & 17.5 & $\text{0.01} \pm \text{0.002}$ & $\text{3.3} \pm \text{0.8}$ \\
PNVC (RA)~\cite{gao2024pnvc}        & -34.87 & -25.74 & -30.24 & -31.98 & -25.86 & -23.03 & 0.01 & 23.6 & 21.8 & 102  \\
NVRC~\cite{kwan2024nvrc}            & -8.52 & -4.92 & -33.59 & -30.62 & -16.13 & -10.66 & $\text{4.4} \pm \text{2.1}$ & $\text{16.5} \pm \text{6.7}$ & $\text{16.8} \pm \text{14.5}$ & $\text{582.1} \pm \text{396.9}$ \\
\midrule
\textbf{GIViC (ours)} & 0.00 & 0.00 & 0.00 & 0.00 & 0.00 & 0.00 & 0.03 & 9.79 & 225.9 & 2399 \\
\bottomrule[1.5pt]
\end{tabular}
}
\caption{Compression performance results of the proposed GIViC framework. Here each BD-rate value is calculated when the corresponding benchmark codec is used as the anchor. Complexity figures for all benchmarked methods have also been provided for comparison.}
\label{tab:bd-psnr}
\end{table*}

\subsection{Optimization}
\noindent \textbf{Loss function.} The above-described per-token generative framework $p(\bmh{x}, \bmh{x}^{0:K}, \bmh{y}^{1:L})$ could be formalized as:
\allowdisplaybreaks
\begin{subequations} \label{eq:decoder}
    \begin{align}
        &p(\bmh{x}^K) \prod\nolimits_{k={K-1}}^{0} 
        \underbrace{p_\theta(\bmh{x}^{k}|\bm{z}^k, \bmh{y}^{1:L})}_{\text{denoising obj.}} 
        \underbrace{p_\psi(\bm{z}^k|\bmh{x}^{K:k+1})}_{\text{recurrent hidden state}} \label{eq:decoder_a} \\
        &p(\bmh{y}^L)
        \underbrace{\prod\nolimits_{l=L-1}^{1} \mathbb{P}_\Phi(\bmh{y}^l|\bmh{y}^{>l})}_{\text{hierarchical prior}} \label{eq:decoder_b}.
    \end{align}
\end{subequations}
This can be re-written using negative log likelihood over all tokens with a GOP-level Lagrange multiplier $\lambda$, yielding the rate-distortion (RD) objective of GIViC:
\begin{equation} \label{eq:rd-loss}
    \mathcal{L}_{\text{RD}} = \mathbb{E}[-\log p(\bm{\mathcal{X}}|\bm{\mathcal{Y}})-\lambda\log p(\bm{\mathcal{Y}})].
\end{equation}

It is noted that when $\bmh{\mathcal{X}}$ is stochastically generated during decoding,  it may result in unstable reconstructions and potentially fail the distortion metric. Given that the majority of open-sourced video compression baselines remain distortion-oriented, we modify the diffusion loss in~\autoref{eq:consistency-obj} using a \textit{post-hoc} guiding mechanism~\cite{dhariwal2021diffusion} to optimize GIViC for the MSE loss.

\vspace{5pt}

\noindent \textbf{Pretraining.} To improve training efficiency, we employ a multi-stage pre-training procedure~\cite{mentzer2022vct,li2023neural}, in which we first fix $(\bm{\mathcal{F}}, \bm{\epsilon})$ of the implicit diffusion and optimize $\mathcal{\bm{E}}_\phi$ and $\mathbb{P}_\Phi$ with a pre-trained quadtree entropy model~\cite{gao2024pnvc} for entropy modeling, which we discard in later stages. We then fix $\mathcal{\bm{E}}_\phi$, swap the HGLA-based backbone with $\mathbb{P}_\Phi$ for the context model, and only update $\mathbb{P}_\Phi$ to optimize the rate loss. Finally, we jointly optimize all components with the RD loss defined by~\autoref{eq:rd-loss}. 

\vspace{5pt}

\noindent \textbf{Encoding.} The latent grids are initialized by running $\mathcal{\bm{E}}_\phi$ and treated as learnable parameters to be iteratively updated during the encoding process. Here we follow~\cite{kim2024c3} to estimate the gradients of $\bmh{\mathcal{Y}}$ based on soft-rounding~\cite{agustsson2020universally} instead of STE, and replace the additive uniform noise with samples from the Kumaraswamy distribution~\cite{kumaraswamy1980generalized} with progressive annealing. Other components in the GIViC framework are fixed during encoding.

\section{Experiment Configuration} \label{sec:experiments}

\noindent \textbf{Implementation.} We pre-trained five baseline models with $\lambda = \{ 85, 170, 380, 840, 1024\}$. By default, the number of diffusive sampling steps is set to 500 at training time and 20 at inference time. The 3D token size is set to $(4, 8, 8)$. Both the MLP-based denoiser $\bm{\epsilon}$ and the HGLA transformer $M$ have a depth of 4 (i.e., $L = M = 4$). All submodules are optimized using the ADAM optimizer~\cite{kingma2014adam} and with an initial learning rate set to $\rm{10}^{-4}$ that is progressively annealed following a cosine scheduling. The models were trained on 4 NVIDIA A100-80G GPUs until convergence.

\vspace{5pt}

\noindent \textbf{Datasets.} GIViC was pretrained on Vimeo-90k~\cite{xue2019video}, and fine-tuned on additional 3,024 videos extracted from raw Vimeo footage, each of which consists of 32 frames, following the practices by~\citep{li2024neural,gao2024pnvc}. For a more comprehensive assessment of GIViC under different training conditions, we have ablated its performance by instead fine-tuning GIViC on the 7-frame Vimeo-90k sequences (V1.1 in~\autoref{tab:ablation}). For testing, we evaluated all models on the UVG~\cite{mercat2020uvg}, MCL-JCV~\cite{wang2016mcl}, and JVET-B~\cite{boyce2018jvet} test sets.

\vspace{5pt}

\noindent \textbf{Baselines.} GIViC is compared against \textit{eight} SOTA baselines, including (i) two \textbf{conventional codecs} - H.265/HEVC Test Model HM 18.0~\cite{hm} and H.266/VVC Test Model VTM 20.0~\cite{vtm}; (ii) three \textbf{neural video codecs} - VCT~\cite{mentzer2022vct}, DCVC-DC~\cite{li2023neural} and DCVC-FM~\cite{li2024neural}, and (iii) three \textbf{INR-based codecs} - C3~\cite{kim2024c3}, PNVC~\cite{gao2024pnvc}, and NVRC~\cite{kwan2024nvrc}. Among these baselines, VTM represents the latest MPEG standard video codec. PNVC and NVRC are the top-performing INR-based codecs. For autoencoder-based methods, while the latest DCVC model, DCVC-LCG~\cite{qi2025long}, has been recently published, its implementation is not yet public. Therefore, we benchmarked against its predecessor, DCVC-FM~\cite{li2024neural}, to represent the state of the art.

\vspace{5pt}

\noindent \textbf{Test conditions.} All experiments on conventional codecs use the Random Access mode defined in JVET common test conditions~\cite{vtm_ctc}. For each rate point, we calculate the bitrate (bit/pixel, bpp) and quantitatively assess the quality of lossy reconstructions using PSNR and MS-SSIM~\cite{wang2003multiscale} in the YUV colorspace. The Bj{\o}ntegaard Delta Rate (BD-rate)~\cite{bdrate} is then used to measure the relative compression efficiency between codecs. We configured PNVC, GIViC, and the conventional codecs in RA mode with a GoPsize=32 and IntraPeriod=32. For HM 18.0 and VTM 20.0, we employed QP = $\{16, 20, 34, 38, 32, 36\}$ to cover a broader range of bitrates. It is important to note that, unlike the other selected baselines, NVRC incurs a system delay equal to the full sequence length and C3 (adaptive) has a system delay corresponding to the patch size, ranging from 30 to 75 frames.

\section{Results and Discussion}

\begin{figure*}[!t]
    \centering
    \includegraphics[width=\textwidth]{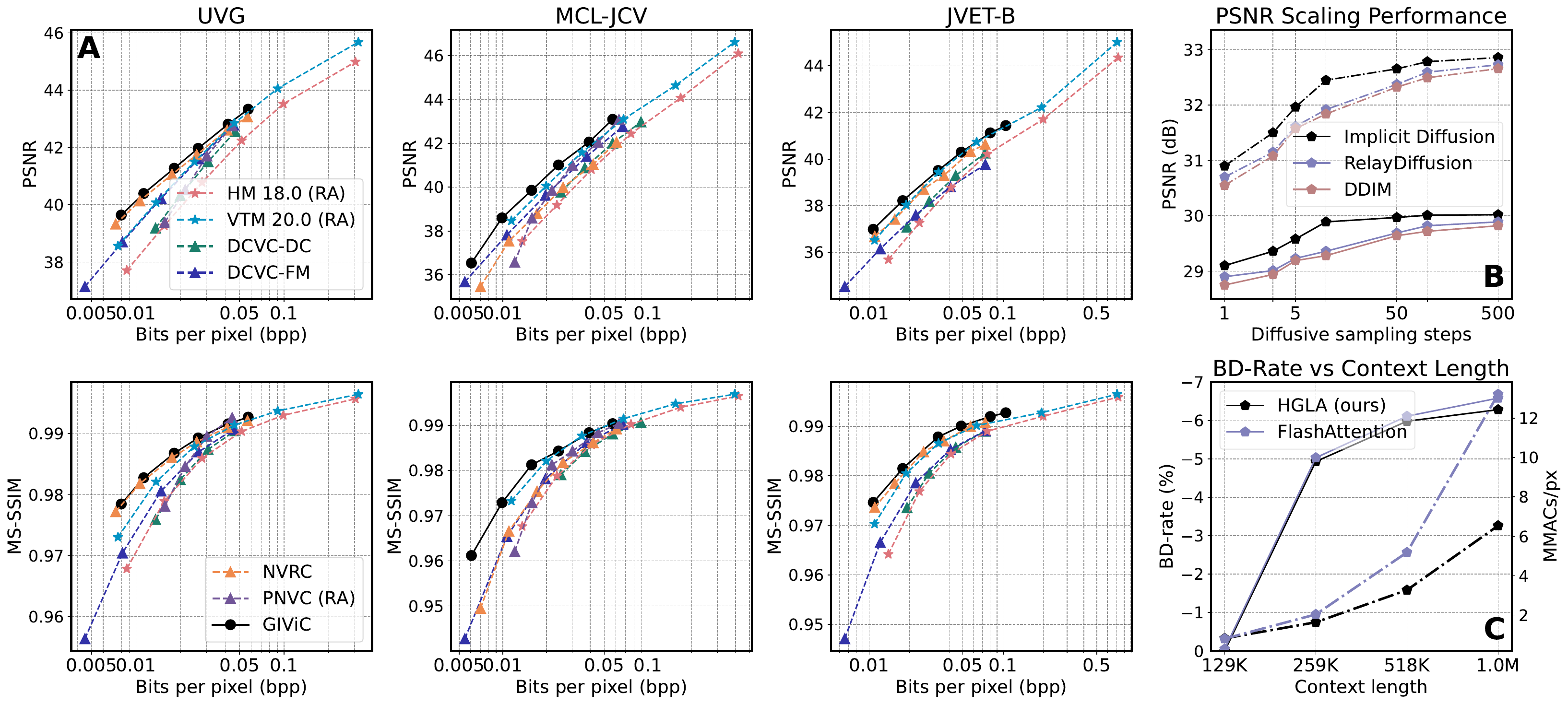}
    \caption{(\textbf{A}) Rate-distortion curves on UVG, MCL-JCV, and JVET-B datasets. (\textbf{B}) Reconstruction quality PSNR w.r.t diffusive sampling steps for low bitrate range (solid lines) and high bitrate range (dashed line) respectively. (\textbf{C}) BD-rate (PSNR, solid lines) and decoding complexity (dashed lines) w.r.t context length.}
    \label{fig:rd}
\end{figure*}

\subsection{Overall Performance}

\noindent \textbf{Quantitative results.} 
The rate-distortion performance of the proposed GIViC codec, compared to conventional and neural video codecs, is summarized in~\autoref{tab:bd-psnr}. Notably, GIViC outperforms all tested neural video codecs, including NVRC, which is regarded as the state-of-the-art INR-based codec. It also surpasses both HM and VTM in terms of coding performance. Specifically, GIViC achieves 15.94\%, 36.68\%, and 8.52\% BD-rate reductions compared to VTM 20.0 (RA), DCVC-FM, and NVRC, respectively. While we cannot directly benchmark GIViC against DCVC-LCG \cite{qi2025long} (the source code of the latter is not available), we have compared GIViC with the results reported in its original literature, which confirms the superior performance of GIViC. \autoref{fig:rd} illustrates the rate-distortion curves of GIViC alongside a selected subset of baseline methods from each category - conventional, autoencoder-based, and INR-based - for three different test sets. The results demonstrate that GIViC maintains consistent performance across diverse datasets throughout the entire tested bitrate range for both $\text{PSNR}$ and $\text{MS-SSIM}$.

\begin{figure*}[t]
    \centering
    \begin{minipage}{0.485\textwidth}
        \centering
        \includegraphics[width=\linewidth]{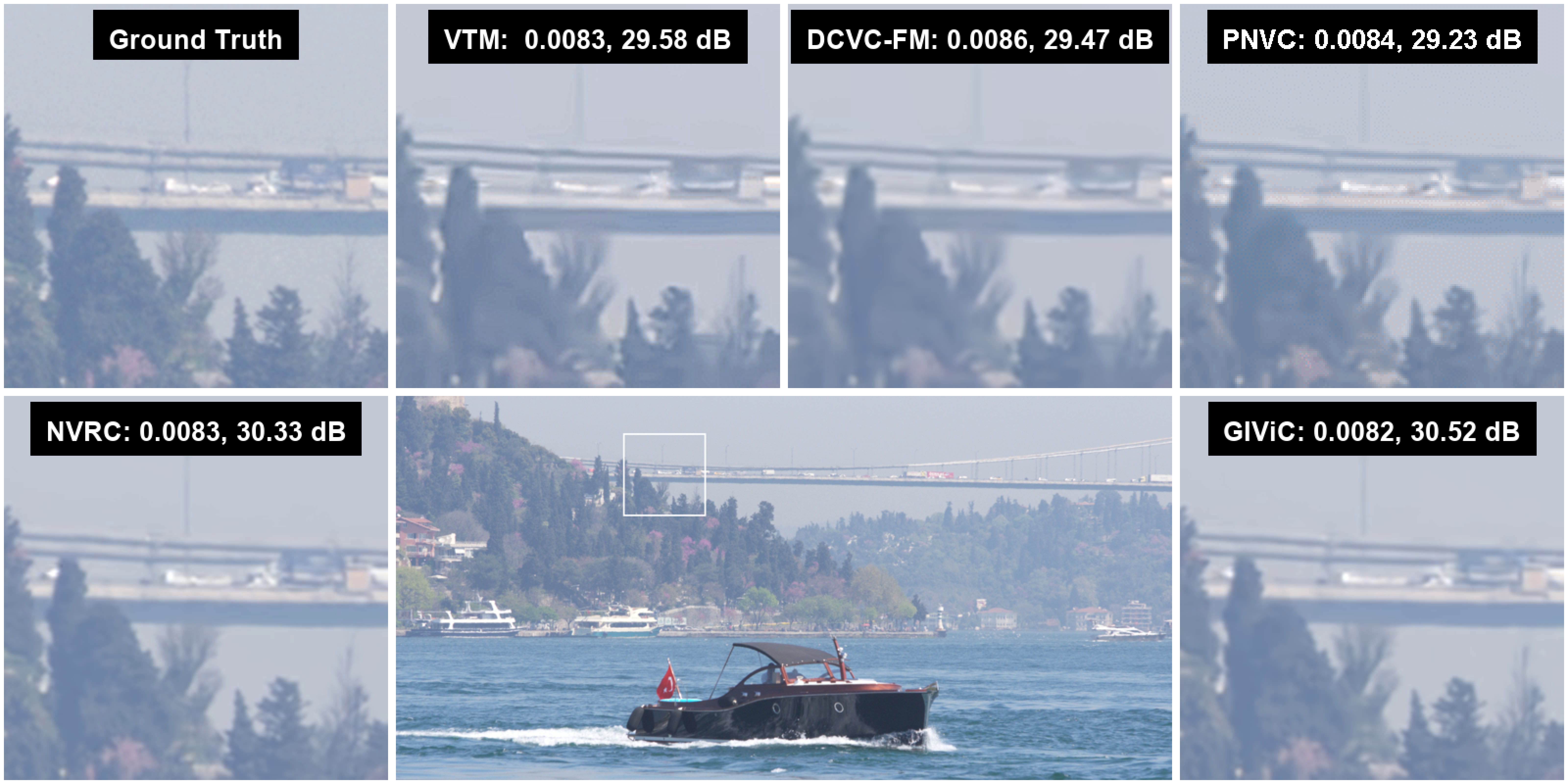}
    \end{minipage}
        \begin{minipage}{0.485\textwidth}
        \centering
        \includegraphics[width=\linewidth]{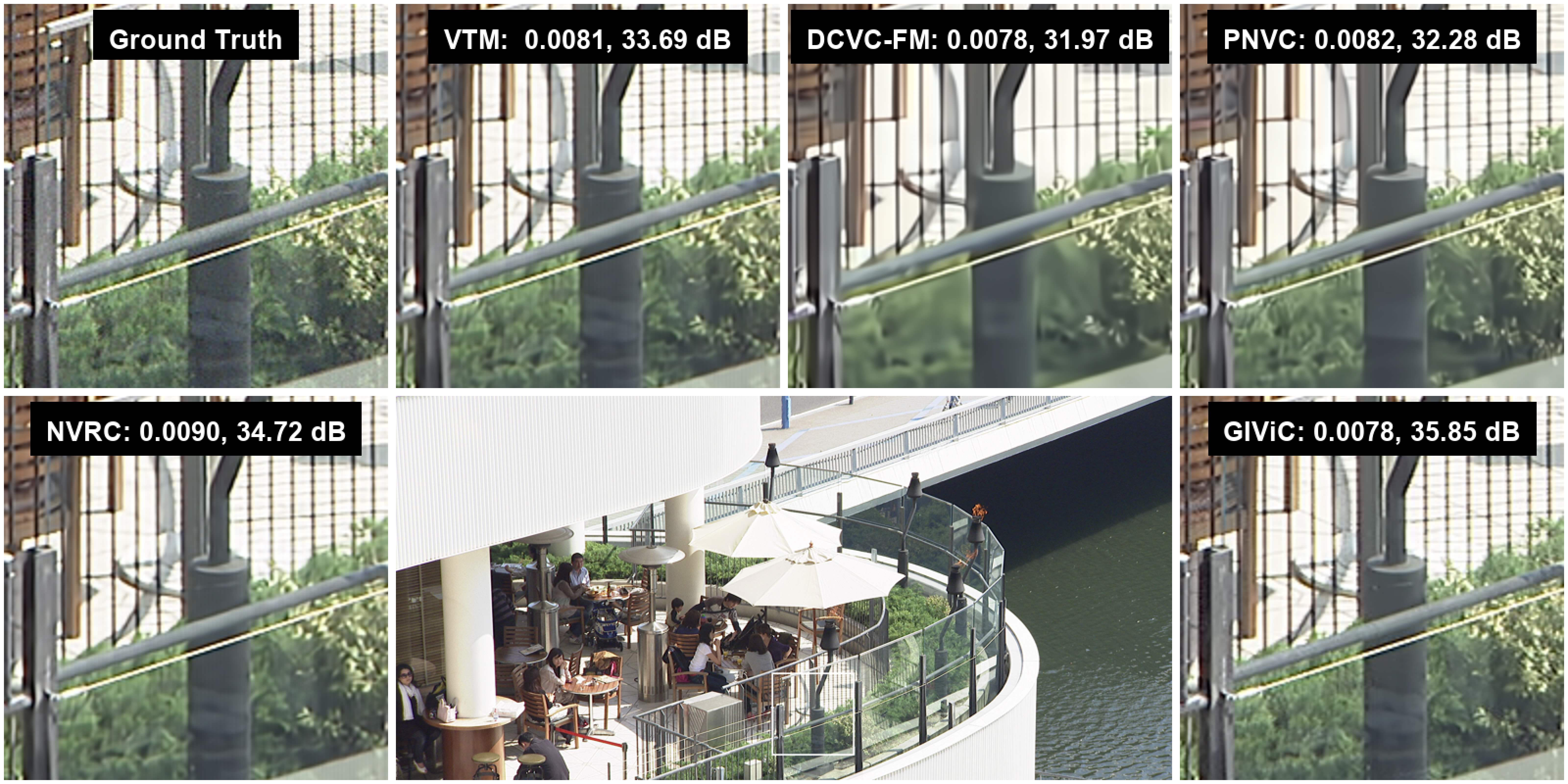}
    \end{minipage}
    \hfill
    \begin{minipage}{0.485\textwidth}
        \centering
        \includegraphics[width=\linewidth]{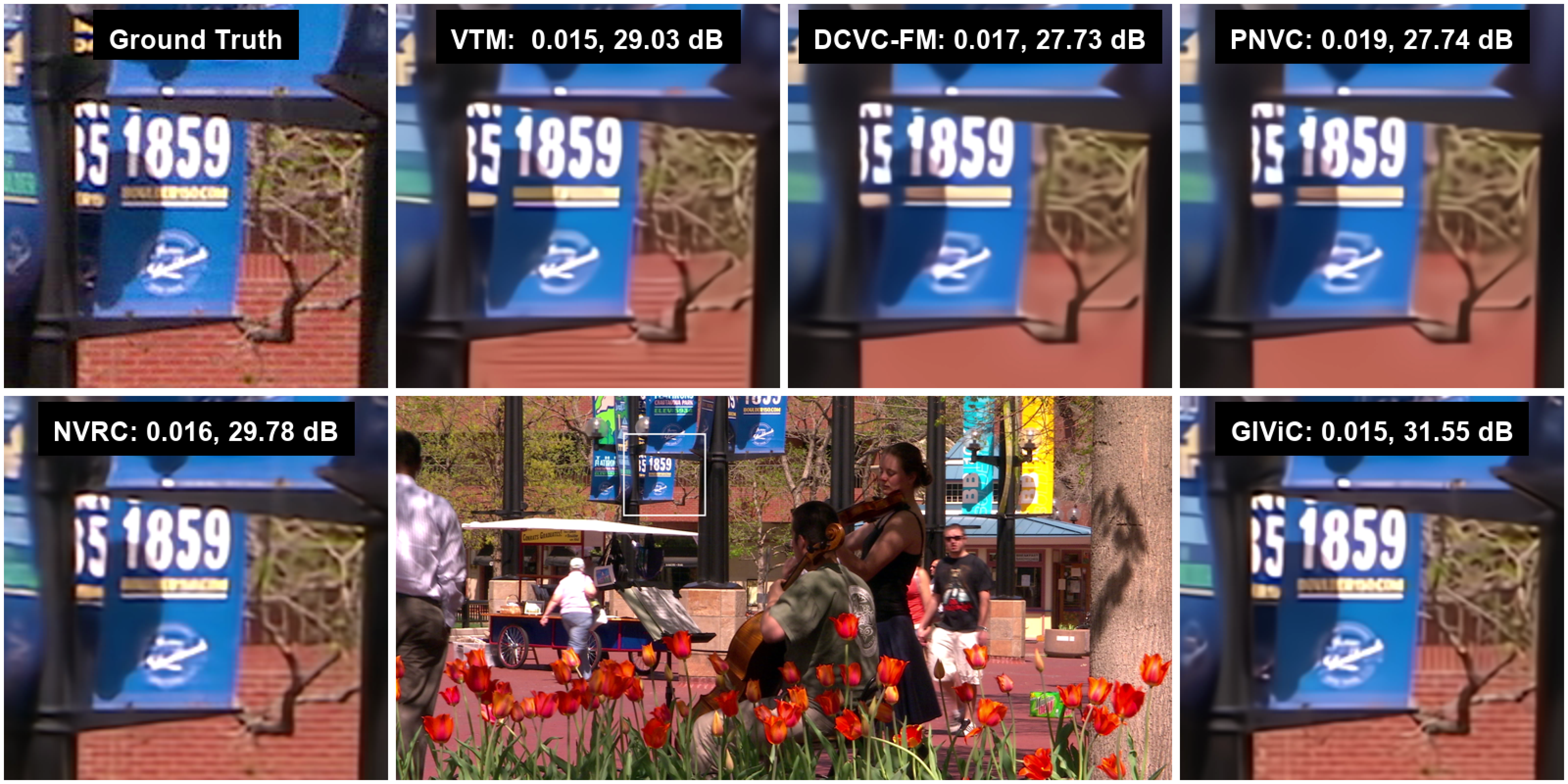}
    \end{minipage}
        \begin{minipage}{0.485\textwidth}
        \centering
        \includegraphics[width=\linewidth]{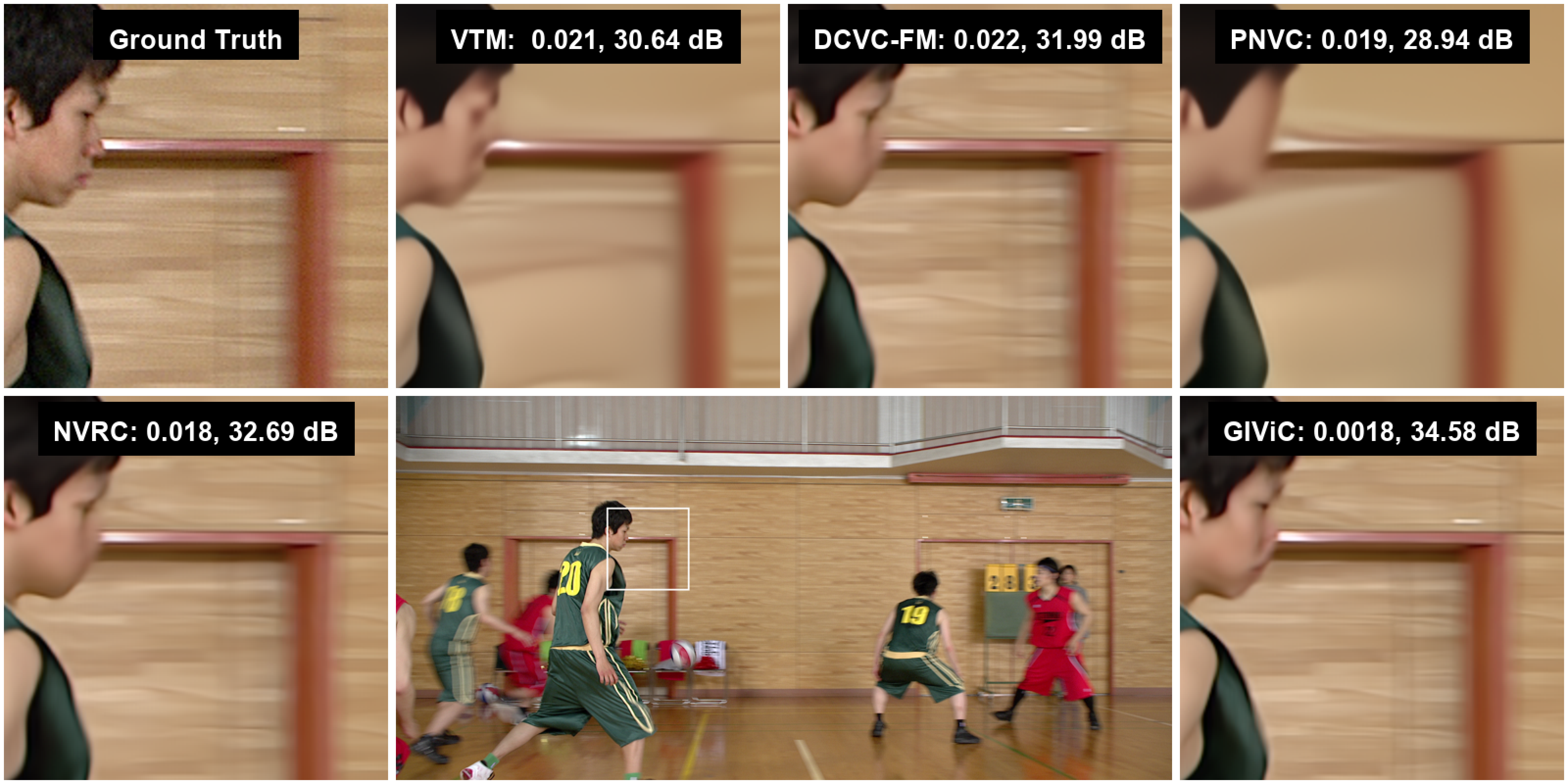}
    \end{minipage}
    
    \caption{Visual comparison of reconstructions by different video codec baselines, where we report the average sequence bpp and the corresponding frame's PSNR.}
    \label{fig:visual-comparison}
\end{figure*}

\vspace{5pt}
\noindent \textbf{Qualitative results.} We further demonstrate the superior performance of our method in terms of subjective visual quality. ~\autoref{fig:visual-comparison} compares frame reconstructions from the proposed GIViC with those from VTM, DCVC-FM, PNVC, and NVRC, showcasing GIViC's improved perceptual quality and reduction of compression artifacts. 

\vspace{5pt}

\noindent \textbf{Complexity.} A complexity profiling is summarized in~\autoref{tab:bd-psnr}. The average encoding and decoding speeds (FPS) for each model are measured on a single NVIDIA A100 GPU. For INR-/overfitting-based methods, the encoding FPS is measured for one full forward and backward pass~\cite{kwan2024hinerv}. It is observed that the training of GIViC takes 2.78 hours per 1000 iterations of optimization on one GOP with 32 frames at $\text{1920} \times \text{1080}$ resolution. The encoding and decoding runtimes of GIViC are, despite being longer, within a range comparable to previously reported INR-based video codecs.

\vspace{5pt}

\noindent \textbf{Scaling performance.} We report the variation in reconstruction or compression performance w.r.t number of diffusive sampling steps and context lengths, in comparison with other diffusion and transformer models. In~\autoref{fig:rd} \textbf{(B)} we compare the average reconstruction fidelity (measured by PSNR) of the proposed implicit diffusion against RelayDiffusion~\cite{teng2023relay} and naive per-token DDIM~\cite{dhariwal2021diffusion} in a single resolution for three lower bitrates (solid lines) and three higher bitrates (dashed lines), respectively. It can be observed that our implicit diffusion scales more favorably and converges at a faster rate for both bitrate ranges. A similar trend is also seen from scaling HGLA on the UVG dataset, shown in~\autoref{fig:rd} \textbf{(C)}, where we configure the GoPSize to 16, 32, 64, and 128, respectively, with the corresponding context length equal to $\frac{\text{GoPSize}\times HW}{r_t \times r_s^2}$. It can be observed that the context length of HGLA leads to a steady, non-trivial, and comparable gain in compression performance with FlashAttention~\cite{dao2022flashattention}, despite its linear instead of quadratic complexity w.r.t the context length.

\subsection{Ablation Study}
We analyze the impact of our methodological contributions and design choices by systematically removing or replacing sub-components and measuring the resulting change in BD-rate and model complexity on two datasets: the UVG and MCL-JCV datasets. The ablative variants include:

\vspace{5pt}
\noindent \textbf{Effectiveness of pre-training} is verified by replacing the 32-frame Vimeo raw sequences with the original Vimeo-90k dataset for fine-tuning (V1.1) and removing the overfitting process in the encoding pipeline (V1.2). 

\vspace{5pt}
\noindent \textbf{Effectiveness of temporal inflation} is tested by instead pretraining a 3D autoencoder entirely from scratch (V2.1) for the same number of optimization steps. 

\vspace{5pt}
\noindent \textbf{Effectiveness of implicit diffusion model} is confirmed by respectively replacing the implicit diffusion with single-resolution per-token diffusion (V3.1), employing RelayDiffusion~\cite{teng2023relay} (V3.2), and replacing the proposed consistency objective with simple flow matching~\cite{liu2022flow} (V3.3). 

\vspace{5pt}
\noindent \textbf{Effectiveness of HGLA} is verified by keeping the original lower bound formulation (V4.1), removing the learned gating across layers (V4.2), and replacing it with a vanilla transformer~\cite{vaswani2017attention} that performs next-scale prediction (V4.3). Further, we ablate the context model $\mathbb{P}_\Phi$ by replacing it with a convolution-based model~\cite{gao2024pnvc} of comparable size (V5.1) and allowing for instance-specific overfitting $\Phi$ following~\cite{gao2024pnvc} (V5.2).

\begin{table}
    \resizebox{\linewidth}{!}{
        \begin{tabular}{r|cc|cc}
            \toprule[1.5pt]
            & \multicolumn{2}{c|}{BD-rate (\%)} & \multicolumn{2}{c}{model complexity} \\
            \cmidrule(lr){2-5}
            version & UVG & MCL & params.(M) & kMACs/px  \\
            \midrule
            \underline{V1.1} & 5.87 & 5.51 & 256.1 (\textcolor{blue}{0.00\%$\downarrow$}) & 2399 (\textcolor{blue}{0.00\%$\downarrow$}) \\
            \underline{V1.2} & 3.66 & 3.19 & 256.1 (\textcolor{blue}{0.00\%$\downarrow$}) & 2399 (\textcolor{blue}{0.00\%$\downarrow$})  \\
            \midrule
            V2.1 & 0.98 & 0.97 & 303.5 (\textcolor{red}{18.6\%$\uparrow$}) & 2806 (\textcolor{red}{16.9\%$\uparrow$}) \\
            \midrule
            V3.1 & 2.35 & 2.69 & 256.1 (\textcolor{blue}{0.00\%$\downarrow$}) & 3590 (\textcolor{red}{49.6\%$\uparrow$}) \\
            \underline{V3.2} & 2.53 & 3.23 & 256.1 (\textcolor{blue}{0.00\%$\downarrow$}) & 2399 (\textcolor{blue}{0.00\%$\downarrow$})  \\
            \underline{V3.3}  & 3.09 & 3.41 & 256.1 (\textcolor{blue}{0.00\%$\downarrow$}) & 2399 (\textcolor{blue}{0.00\%$\downarrow$})  \\
            \midrule
            V4.1 & 1.99 & 2.43 & 256.1 (\textcolor{blue}{0.00\%$\downarrow$}) & 2399 (\textcolor{blue}{0.00\%$\downarrow$}) \\
            V4.2 & 3.17 & 3.20 & 223.8 (\textcolor{blue}{0.09\%$\downarrow$}) & 2371 (\textcolor{blue}{1.27\%$\downarrow$}) \\
            V4.3 & 0.19 & 0.25 & 219.7 (\textcolor{blue}{2.74\%$\downarrow$}) & 2987 (\textcolor{red}{24.5\%$\uparrow$}) \\
            \midrule
            V5.1 & 6.77 & 6.25 & 221.2 (\textcolor{blue}{2.17\%$\downarrow$}) & 2380 (\textcolor{blue}{0.79\%$\downarrow$}) \\
            V5.2 & 3.20 & 2.65 & 221.2 (\textcolor{blue}{2.17\%$\downarrow$}) & 2380 (\textcolor{blue}{0.79\%$\downarrow$}) \\
            \bottomrule[1.5pt]
        \end{tabular}
    }
    \captionof{table}{Ablation study results on the UVG and MCL-JCV datasets in terms of BD-rates (measured in PSNR), and the entire model's size and kMACs/pixel, measured against GIViC. Here, the \underline{underlined} ablative variants are irrelevant to architectural modifications and thus incur no changes in model complexity.}
    \label{tab:ablation}
\end{table}

The ablation study results are summarized in~\autoref{tab:ablation}, from which it can be observed that all these ablative variants result in compression loss when compared to the original GIViC. This indicates that each contribution in this work does improve the overall compression performance.

\section{Conclusion} \label{sec:conclusion}
This paper proposes GIViC, an INR-based video coding framework that integrates generalized diffusion and linear attention-based transformers for advanced video compression. The implicit diffusion process leverages continuous spatiotemporal downsampling, paired with a cascaded denoising process stitching multiple spatiotemporal resolutions. In addition, we introduce HGLA, a novel transformer architecture for long-term dependency modeling that captures both cross-scale and bidirectional spatiotemporal dependencies at linear complexity. GIViC achieves superior compression performance over SOTA video codecs including VTM 20.0 (by 15.94\%, 11.44\%, and 7.71\%) and NVRC (by 8.52\%, 33.59\%, and 16.13\%) under the RA configuration on UVG, MCL-JCV, and JVET-B datasets, respectively. As far as we are aware, this is the \textbf{first time that an INR-based video codec has outperformed VTM with the same latency constraint (RA)}. However, it is also noted that GIViC does exhibit relatively high computational complexity due to its reliance on diffusion and transformer backbones, making its adoption more challenging for applications that require real-time decoding speeds. This issue remains a topic for future investigation.

{\small
\bibliographystyle{ieeenat_fullname}
\bibliography{main}}

\end{document}